\documentclass[a4paper,11pt]{article}   	
\usepackage{geometry}                		
\usepackage{amsmath}
\usepackage{amssymb}
\usepackage{graphicx}
\usepackage{mathrsfs}
\usepackage{hyperref}
\usepackage{braket}
\usepackage{cite}
\usepackage{authblk}
\usepackage{xcolor}

\def\eref#1{(\ref{eq:#1})}

\def\squarebr#1{\left[#1\right]}

\def\del#1#2{\frac{\partial #1}{\partial #2}} 
\def\D{\mathrm{d}}

\def\I{\mathrm{i}}
\def\Ob{\mathcal{O}}

\def\He{\hat{H}_{\text{eff}}}
\def\Hs{\mathsf{H}}
\def\W{\mathrm{W}}
\def\Stt{\mathtt{S}}
\newcommand{\biggpar}[1]{\bigg( #1 \bigg)}
\newcommand{\Om}{\Ob\left(\frac{1}{M}\right)}
\newcommand{\Omm}{\Ob\left(\frac{1}{M^2}\right)}

\def\eq#1{\begin{equation}\begin{aligned}#1\end{aligned}\end{equation}}

\usepackage[normalem]{ulem}

\usepackage{xcolor}

\begin{document}

\title{Beyond semiclassical time}
 
\author{Leonardo Chataignier\thanks{leonardo.chataignier@unibo.it}}
\affil{Dipartimento di Fisica e Astronomia, Universit\`{a} di Bologna,
via Irnerio 46, 40126 Bologna, Italy}
\affil{I.N.F.N., Sezione di Bologna, I.S. FLAG, viale B. Pichat 6/2, 40127 Bologna, Italy} 
 
\date{}

\maketitle
\begin{abstract}
We show that the usual Born-Oppenheimer type of approximation used in quantum gravity, in which a semiclassical time parameter emerges from a weak-coupling expansion of the Wheeler-DeWitt constraint, leads to a unitary theory at least up to the next-to-leading order in minisuperspace models. As there are no unitarity-violating terms, this settles the issue of unitarity at this order, which has been much debated in the literature. Furthermore, we also show that the conserved inner product is gauge-fixed in the sense that the measure is related to the Faddeev-Popov determinant associated with the choice of semiclassical time as a reparametrization gauge. This implies that the Born-Oppenheimer approach to the problem of time is, in fact, an instance of a relational quantum theory, in which transition amplitudes can be related to conditional probabilities.
\end{abstract}\pagebreak

\section{Introduction}
One of the many challenges that must be overcome before a complete theory of quantum gravity is at hand is the well-known ``problem of time''~\cite{Kuchar:1991,Isham:1992,Kiefer:book,Anderson:book}, which is summarized by the questions: (1) Can the quantum-gravitational dynamics be described with respect to a certain (or several) time variable(s)? (2) How can we define transition probabilities? (3) Is the quantum-gravitational dynamics unitary, i.e., are probabilities conserved? One sees that, should quantum gravity follow the standard structure of ordinary quantum theories, it would not only be necessary to specify a choice of time/dynamics, but also a conserved inner product. The absence of an obvious choice of time stems from the Wheeler-DeWitt (WDW) constraint~\cite{DeWitt:1967}
\eq{\label{eq:WDW}
\hat{H}\Psi = 0 \ , 
}
which is a time-independent Schr\"{o}dinger equation for the wave function(al) of matter and gravitational fields that follows from the diffeomorphism invariance of the theory~\cite{Kiefer:book}.

Several different approaches have been devised to extract a meaningful notion of dynamics from~\eref{WDW}. Noteworthy are the `Born-Oppenheimer' (BO)~\cite{LapRuba:1979,Banks:1984,Halliwell:1984,Brout:1987-4,PadSingh:1990-1,PadSingh:1990-2,Singh:1990,Kiefer:1991,Bertoni:1996,Kiefer:1993-2,Parentani:1997-2,Bologna:2017,Kiefer:2018,Chataig:2019-1} and the `relational observables' approaches~\cite{Rovelli:1990-1,Rovelli:1990-2,Rovelli:1991,Dittrich:2004,Dittrich:2005,Tambor,Chataig:2019-2,Hoehn:2018-1,Hoehn:2018-2,Hoehn:2019,Hoehn:Trinity,Chataig:2020,Hoehn:2020,Chataig:Thesis}. The former approach was inspired by the Born-Oppenheimer approximation scheme in molecular physics~\cite{BO:1927,Cederbaum:2008,Abedi:2010,Arce:2012}, which is well suited for models in which the dynamical fields can be separated according to characteristic energy scales $m\ll\sqrt{M}$, such that a formal perturbative expansion of~\eref{WDW} in powers of $m^2/M$ or, more schematically, $1/M$ can be performed. In quantum gravity, $m$ can be a generic matter energy scale, whereas $M$ usually corresponds to the square of the Planck mass. In this way, the series in powers of $1/M$ corresponds to a `weak-coupling expansion'~\cite{Chataig:2019-1}. The key aspect is that a time parameter `emerges' from this expansion, and it can be interpreted as the `orderer' of the lowest-order (semiclassical) dynamics of gravitation, which neglects the backreaction of matter fields. On the other hand, in the approach of relational observables, the evolution is understood in relation to a reference field, the level sets of which define the instants of time. This strategy can in principle be defined beyond the semiclassical level; i.e., it is applicable without resorting to a perturbative scheme, and a meaningful notion of evolution is available even if all the fields are quantum and interacting.

In the present article, we show that the BO approach is, in fact, an instance of the relational strategy, which is more general. Concretely, we focus on minisuperspace theories that correspond to general mechanical BO systems, and we show how the weak-coupling expansion is connected to a gauge-fixed inner product, where the choice of gauge corresponds to the definition of semiclassical time, thus fixing the freedom associated with diffeomorphism and general coordinate invariance. The `measure' in the inner product corresponds to a quantization of the absolute value of the classical Faddeev-Popov (FP) determinant~\cite{FP-1,FP-2,HT:book} connected to the gauge choice. At least up to the next-to-leading order (NLO), this inner product is conserved with respect to the semiclassical time, thus settling the issue of unitarity of the BO approach, which has been a much debated topic in the literature~\cite{Bertoni:1996,Kiefer:1991,Kiefer:2018,Chataig:2019-1,BKK-1,BKK-2,Bologna:2017}. Furthermore, relational observables can be defined from their matrix elements by suitable operator insertions into the gauge-fixed inner product. In this way, the BO scheme can be seen as a perturbative approach to the relational quantum dynamics, and its importance lies in the practicality of perturbation theory, which renders phenomenological calculations, such as corrections to the power spectrum of primordial fluctuations~\cite{BKK-0-0,BKK-0-1,Bologna:2013,Bologna:2014,Bologna:2016,Bologna:2018,Stein1,Stein2,Mariam,BKK-3,Bologna:2020,BKK-1,BKK-2,Bologna:2017,Chataig:2021}, possible. The notion of time, however, is relationally meaningful beyond the semiclassical level.

This formalism is an extension to general mechanical models of what was found in~\cite{Chataig:2021} for the particular case of a quasi-de Sitter universe with cosmological perturbations, which in turn was a continuation of the work done in~\cite{Chataig:2019-1}. Moreover, the results presented here were previously reported in the author's thesis~\cite{Chataig:Thesis}, to which the reader is referred for further details. Throughout the article, we adopt units in which $c=\hbar=1$ and we omit the summation sign over repeated indices.

\section{Classical theory}
Before we discuss the corresponding quantum formalism, it is useful to consider how a weak-coupling expansion of the classical constraint leads to a ``preferred'' choice of gauge. Let the action be
\begin{equation}\label{eq:action}
S = \int\mathrm{d}\tau\ \left(P_a\dot{Q}^a+p_{\mu}\dot{q}^{\mu}-N(\tau)H\right) \ ,
\end{equation}
where $\cdot\equiv\D/\D\tau$ and $a = 1,\ldots,n$, $\mu = 1,\ldots,d$. The Hamiltonian constraint $H=0$ is obtained by varying~\eref{action} with respect to the lapse $N(\tau)$, which plays the role of a Lagrange multiplier. We assume that $H$ is of the form
\eq{\label{eq:Hconstraint}
H = \frac{1}{2M}G^{ab}(Q)P_aP_b+MV(Q)+\mathsf{H}(Q;p,q) \ ,
}
where $V(Q)$ is assumed to be non-vanishing and $\Hs(Q;p,q)$ is a non-negative smooth function of zeroth order in the scale $M$. The action~\eref{action} is invariant under time reparametrizations $\tau\to\tau'$ because the lapse transforms as $N(\tau)\D\tau = N'(\tau')\D\tau'$ and the other variables are reparametrization scalars. A fixation of the arbitrary form of the lapse corresponds to a choice of gauge.

Finding the trajectories associated with~\eref{Hconstraint} might be complicated in practice and some perturbative expansion is warranted. To this end, let us consider the Hamilton-Jacobi (HJ) equation associated with~\eref{Hconstraint},
\begin{equation}\label{eq:HJ-general}
\frac{1}{2M}G^{ab}(Q)\frac{\partial\W}{\partial Q^a}\frac{\partial\W}{\partial Q^b}+MV(Q)+\Hs\left(Q;\frac{\partial \W}{\partial q},q\right) = 0 \ ,
\end{equation}
where $\W$ is Hamilton's characteristic function. Since $-1$ is the lowest power of $1/M$ in~\eref{HJ-general}, we can consider a type of Wentzel-Kramers-Brillouin (WKB) ansatz for $\W$,
\eq{\label{eq:pertW}
\W(Q,q) = M\sum_{n = 0}^{\infty}\W_n(Q,q)\frac{1}{M^n} =: M\W_0(Q)+\mathtt{S}(Q;q) \ ,
}
and solve~\eref{HJ-general} order by order; i.e., the weak-coupling expansion consists in solving for all the $\W_n$ functions. In particular, we find that $\W_0$ must be a solution to
\eq{\label{eq:background-HJ}
\frac{1}{2}G^{ab}(Q)\frac{\partial\W_0}{\partial Q^a}\frac{\partial\W_0}{\partial Q^b}+V(Q) = 0 \ ,
}
which corresponds to the HJ constraint for the $Q$ variables without the backreaction of the $q$ fields. This is the no-coupling limit.

\subsection{Dynamics in the no-coupling limit}\label{sec:no-coupling}
It is useful to define a set of coordinates in the configuration space which is `adapted' to the Hamilton function $\W_0$~\cite{Chataig:2019-1}. To this end, we consider a function $x^1(Q)$ which solves~\cite{Chataig:2019-1,Kuchar:1991,Isham:1992}
\eq{
G^{ab}\del{\W_0}{Q^a}\del{x^1}{Q^b} = 1 \ ,
}
such that its equation of motion in the no-coupling limit reads [cf.~\eref{action}]
\eq{
\dot{x}^1(Q) = \del{x^1}{Q^a}\dot{Q}^a =\del{x^1}{Q^a}N(\tau)G^{ab}\del{\W_0}{Q^b} = N(\tau) \ .
}
Thus, $x^1(q)$ corresponds to proper time when the coupling with the $q$ fields is neglected. It is also frequently called `WKB time'~\cite{Zeh} because it originates in the WKB expansion~\eref{pertW}. This also leads to the identity
\eq{\label{eq:dQadx1}
\del{Q^a}{x^1} \equiv \left.\dot{Q}^a\right|_{N(\tau)=1} = G^{ab}\del{\W_0}{Q^b} \ .
}
For simplicity, we then assume that it is possible to foliate the $Q$-configuration space according to the level sets of the function $x^1(Q)$ and that we can define coordinates $x(Q) = (x^1(Q),x^i(Q))$ with respect to which the metric components read [cf.~\eref{background-HJ} and~\eref{dQadx1}]
\eq{\label{eq:normalization-basis-vectors}
\tilde{G}_{11} &= G_{ab}\del{Q^a}{x^1}\del{Q^b}{x^1} = -2V(Q(x)) \ , \\
\tilde{G}_{1i} &= 0 \ , \\
\tilde{G}_{ij} &=: g_{ij} \ ,\\
g &:= \det\mathbf{g} \ , \\
\tilde{G} &:=\det\tilde{\mathbf{G}} =  -2Vg\ ,
}
for $i,j = 2, \ldots, n$. The choice of coordinates~\eref{normalization-basis-vectors} also implies [cf.~\eref{dQadx1}]
\begin{equation}\label{eq:x-gradient-varphi}
\begin{aligned}
\frac{\partial\W_0}{\partial x^1} &= -2V(Q(x)) \ , \\
\frac{\partial\W_0}{\partial x^i} &= 0 \ ,
\end{aligned}
\end{equation}
and, consequently,
\eq{\label{eq:nabla2W0}
\nabla^2\W_0 &= \frac{1}{\sqrt{|\tilde{G}|}}\del{}{x^1}\left(\tilde{G}^{11}\sqrt{|\tilde{G}|}\del{\W_0}{x^1}\right)\\
&=\del{}{x^1}\log\sqrt{|2Vg|} \ ,
}
which will be useful in the quantum theory.

\subsection{WKB time is a choice of gauge}
By taking higher powers of $1/M$ into account, the effect of the coupling between the $Q$ and $q$ fields becomes non-negligible. Since the higher orders in the weak-coupling expansion are encoded in the function $\Stt$ given in~\eref{pertW}, it is this function that captures the coupled dynamics. In fact, $\Stt$ describes the canonical system when the effect of the no-coupling limit is subtracted. This corresponds to the canonical transformation
\begin{equation}\label{eq:background-canonical}
\begin{aligned}
\del{Q^a}{x^A}P_a = \frac{\partial \W}{\partial x^A}&\mapsto\Pi_A = \del{Q^a}{x^A}P_a - M\frac{\partial\W_0}{\partial x^A} = \frac{\partial \Stt}{\partial x^A} \ ,\\
p_{\mu} = \frac{\partial \W}{\partial q^{\mu}}&\mapsto p_{\mu} = \frac{\partial \Stt}{\partial q^{\mu}} \ ,
\end{aligned}
\end{equation}
where $A = 1,\ldots,n$. It is straightforward to see that the Lagrangian of the canonical system described by~\eref{background-canonical} differs from the one in~\eref{action} by a total time derivative. If the characteristic Hamilton function $\W$ solves~\eref{HJ-general}, then $\Stt$ must be a solution to
\eq{\label{eq:HJ-S}
\del{\Stt}{x^1}+\Hs+\frac{1}{2M}g^{ij}\del{\Stt}{x^i}\del{\Stt}{x^j}-\frac{1}{4M V}\left(\del{\Stt}{x^1}\right)^2 = 0 \ ,
}
where we used the coordinate transformation~\eref{normalization-basis-vectors}. Even though the dynamics of $Q$ and $q$ is coupled, we may still wish to parametrize it by the WKB time $x^1$, which corresponds to marking the passage of time according to a `background clock' that neglects the effects of the coupling. The evolution with respect to this clock is dictated by the `reduced Hamiltonian' that corresponds to the opposite of its momentum,
\eq{\label{eq:physical-H}
-\Pi_1 := -\del{\Stt}{x^1} = -2M V\pm2 M\sqrt{V\left(V+\frac{1}{M}\Hs+\frac{1}{2M^2}g^{ij}\del{\Stt}{x^i}\del{\Stt}{x^j}\right)} \ .
}
The second line of the above equation follows from~\eref{HJ-S}. This choice of time parameter fixes the form of the lapse function $N(\tau)$ by the requirement [cf.~\eref{HJ-general},~\eref{normalization-basis-vectors},~\eref{background-canonical} and~\eref{HJ-S}]
\eq{\label{eq:lapse}
1 = \dot{x}^1 = \frac{N(\tau)}{M}\tilde{G}^{11}\del{\W}{x^1} = N(\tau)\left(1-\frac{\Pi_1}{2MV}\right) \ .
}
Furthermore, the FP determinant $\Delta$~\cite{FP-1,FP-2} is defined to be the determinant of the matrix of Poisson brackets between gauge conditions and constraints. As there is only one constraint [cf.~\eref{Hconstraint}] and one gauge condition $x^1(\tau) = \tau$ [cf.~\eref{lapse}], this becomes
\eq{\label{eq:FPdet}
\Delta :=\det\{x^1,H\} = \frac{1}{N(\tau)}\dot{x}^1 = 1-\frac{\Pi_1}{2MV} \ .
}

\subsection{Perturbation theory}
It is possible and, in fact, more tractable to expand~\eref{physical-H} and~\eref{FPdet} in powers of $1/M$ and continue with the perturbative analysis initiated in Sec.~\ref{sec:no-coupling}. Up to the first non-trivial order, we find
\begin{equation}\label{eq:physical-H-expanded}
\begin{aligned}
-\Pi_1 = -2MV+\sigma\biggpar{2M|V|+\mathfrak{v}\Hs-\frac{1}{4M|V|}\Hs^2+\frac{\mathfrak{v}}{2M}g^{ij}\Pi_i \Pi_j}+\mathcal{O}\left(\frac{1}{M^2}\right) \ ,
\end{aligned}
\end{equation}
and
\eq{\label{eq:gauge-fixed-lapse-expanded}
\Delta = \sigma\mathfrak{v}+\frac{\sigma\Hs}{2M|V|}+\Omm \ ,
}
where $\sigma=\pm1$ and $\mathfrak{v}:=\mathrm{sgn}(V)$. For the quantum theory, it will be useful to consider the case in which $\sigma =\mathfrak{v}$, which leads to the simplified formulae
\begin{equation}\label{eq:Hamiltonian-BO}
-\Pi_1 = \Hs-\frac{1}{4MV}\Hs^2+\frac{1}{2M}g^{ij}\Pi_i\Pi_j+\mathcal{O}\left(\frac{1}{M^2}\right) \ ,
\end{equation}
and
\eq{\label{eq:absolute-FP}
|\Delta| = 1+\frac{\Hs}{2MV}+\Omm \ .
}
This case $\sigma =\mathfrak{v}$ is what is obtained if one solves~\eref{HJ-S} for $\Pi_1$ in an iterative way and it is analogous to what will be done in the quantum theory, where the quantum constraint will be solved iteratively.

\section{Quantum theory}
The canonical quantization of the Hamiltonian constraint~\eref{Hconstraint} with Laplace-Beltrami factor ordering leads to the time-independent Schr\"{o}dinger equation
\eq{\label{eq:WDW-BO}
0 = \hat{H}\Psi = -\frac{1}{2M}\nabla^2\Psi+MV(Q)\Psi+\hat{\Hs}\Psi \,,
}
which is the Wheeler-DeWitt constraint~\eref{WDW}. The operator $\nabla^2$ reads\footnote{\label{foot:h}If $\hat{\Hs}$ involves a non-trivial metric for the $q$-configuration space that also depends parametrically on $Q$, then the determinant $G$ in~\eref{nabla2} may be replaced by $G\to Gh$, where $h$ is the determinant of the $q$-sector metric~\cite{Chataig:2019-1,Chataig:Thesis}.}
\eq{\label{eq:nabla2}
\nabla^2=\frac{1}{\sqrt{|G|}}\frac{\partial}{\partial Q^a}\left(\sqrt{|G|}G^{ab}\frac{\partial}{\partial Q^b}\right) \ ,
}
and it is invariant under coordinate transformations in configuration space.

\subsection{Phase transformation of the quantum constraint}
In analogy to~\eref{pertW}, we assume that the wave function can be expanded according to
\eq{\label{eq:BO-ansatz}
\Psi(Q,q) = \exp\left[\I M \sum_{n = 0}^{\infty}\mathcal{W}_n(Q,q)\frac{1}{M^n}\right] =: \exp\squarebr{\I M \mathcal{W}_0(Q,q)}\psi(Q;q) \ .
}
As the lowest power of $1/M$ in~\eref{WDW-BO} appears together with $V(Q)$, we assume for simplicity that $\mathcal{W}_0(Q,q)\equiv\mathcal{W}_0(Q)$. In this case, the expansion~\eref{BO-ansatz} corresponds to the BO ansatz.\footnote{\label{foot:traditional}In~\cite{Chataig:2021,Chataig:Thesis} it is explained that this expansion of the wave function is indeed equivalent to the traditional BO factorization $\Psi = \psi_0(Q)\psi(Q;q)$. Moreover, the justification of a single phase factor multiplying $\psi(Q;q)$ in~\eref{BO-ansatz} instead of a superposition of the type $\exp\left(\I M \mathcal{W}_0(Q)\right)+\exp\left(-\I M \mathcal{W}_0(Q)\right)$ may be a consequence of decoherence~\cite{Kiefer:topology}.} 

At the lowest order, we then find that $\mathcal{W}_0(Q)$ must be a solution to~\eref{background-HJ}, such that we may simply impose $\mathcal{W}_0(Q) = \W_0(Q)$. Using~\eref{BO-ansatz} as a phase transformation [where the phase is $\W_0(Q)$], we see that if $\Psi$ solves~\eref{WDW-BO}, then $\psi(Q;q)$ obeys
\begin{equation}\label{eq:pre-schrodinger}
\I G^{ab}\del{\W_0}{Q^a}\frac{\partial \psi}{\partial Q^b} = \left[\hat{\Hs}-\frac{\I}{2}\nabla^2\W_0\right]\psi-\frac{1}{2M}\nabla^2\psi \ ,
\end{equation}
which is the phase-transformed constraint. If we now use the coordinate transformation~\eref{normalization-basis-vectors} together with~\eref{nabla2W0}, we can rewrite~\eref{pre-schrodinger} as
\eq{\label{eq:pre-schrodinger-2}
\I\del{}{x^1}\left(|2Vg|^{\frac14}\psi\right) = |2Vg|^{\frac14}\hat{\Hs}\psi-\frac{|2Vg|^{\frac14}}{2M}\nabla^2\psi \ .
}
Since $\hat{\Hs}\equiv\hat{\Hs}(Q;\hat{p},q)$, we find $|2Vg|^{\frac14}\hat{\Hs} = \hat{\Hs}|2Vg|^{\frac14}$.\footnote{Following footnote~\ref{foot:h}, if the $q$-sector metric has a non-trivial determinant $h(Q,q)$, then a factor such as $|h|^{\frac14}$ might not commute with $\hat{\Hs}$ depending on the choice of factor ordering. See~\cite{Chataig:2019-1,Chataig:Thesis} for a more general treatment.} Furthermore, by expressing the Laplace-Beltrami operator $\nabla^2$ with respect to the $x$ coordinates, we can rearrange~\eref{pre-schrodinger-2} into
\eq{\label{eq:quasi-schrodinger}
&\I\del{}{x^1}\left(|2Vg|^{\frac14}\hat{\mu}\psi\right) = \hat{\mathfrak{H}}\psi\ ,
}
where we defined
\eq{\label{eq:pre-FP}
\hat{\mu} &:= 1+\frac{\I}{4MV}|2Vg|^{-\frac14}\del{}{x^1}|2Vg|^{\frac14} \ ,
}
and
\eq{\label{eq:pre-Heff}
\hat{\mathfrak{H}}\psi&:=\hat{\Hs}|2Vg|^{\frac14}\psi-\frac{|2Vg|^{-\frac14}}{2M}\del{}{x^i}\left(\sqrt{|2Vg|}g^{ij}\del{\psi}{x^j}\right)\\
&\ \ +\del{}{x^1}\left[\del{}{x^1}\left(|2Vg|^{-\frac14}\right)\frac{\sqrt{|2Vg|}}{4MV}\right]\psi \ .
}
Notice that the operator that corresponds to the momentum conjugate to $x^A$ ($A=1,\ldots,n$) relative to the metric given in~\eref{normalization-basis-vectors} is~\cite{DeWitt:1957}
\eq{\label{eq:momentum-operator}
\hat{\Pi}_A := -\I |2Vg|^{-\frac14}\del{}{x^A}|2Vg|^{\frac14} \ \ (A=1,\ldots,n) \ ,
}
such that~\eref{pre-FP} is equivalent to
\eq{\label{eq:pre-FP-2}
\hat{\mu} &:= 1-\frac{1}{4MV}\hat{\Pi}_1 \ .
}

\subsection{Perturbation theory}
Equation~\eref{quasi-schrodinger} can be solved iteratively using perturbation theory. Up to the lowest order in $1/M$, we can use~\eref{quasi-schrodinger} together with~\eref{pre-Heff} to obtain
\eq{\label{eq:schrodinger}
\I\del{\tilde{\psi}}{x^1} = \hat{\Hs}\tilde{\psi}+\Om \ ,
} 
where we defined
\eq{\label{eq:conditional-wave}
\tilde{\psi}:=|2Vg|^{\frac14}\psi+\Om \ .
}
Equation~\eref{schrodinger} is simply the Schr\"{o}dinger equation for the $q$ variables, where the $x$ coordinates behave as parameters or c-numbers. Due to~\eref{dQadx1} and~\eref{normalization-basis-vectors}, these parameters follow the classical trajectory for the $Q$ fields determined by $\W_0$. In this sense, the dynamics at the lowest order is semiclassical: the $q$ variables evolve according to the quantum equation~\eref{schrodinger}, whereas the $Q(x)$ fields remain classical. For this reason, WKB time is also sometimes referred to as `semiclassical' time~\cite{Kiefer:1993-2}. Equation~\eref{schrodinger} illustrates how the weak-coupling expansion in powers of $1/M$ allows one to recover the standard Schr\"{o}dinger-picture of the quantum (field) theory for the $q$ fields from the (time-independent) Hamiltonian constraint~\cite{LapRuba:1979,Banks:1984,Halliwell:1984,Brout:1987-4,PadSingh:1990-1,PadSingh:1990-2,Singh:1990,Kiefer:1991,Bertoni:1996,Kiefer:1993-2,Chataig:2019-1}.

Moreover, up to the NLO, we can use~\eref{pre-FP-2} to write
\eq{\label{eq:qFP}
\hat{\mu}^2 = 1-\frac{1}{2MV}\hat{\Pi}_1+\Omm \ ,
}
which is seen to be a quantization (with a particular factor ordering) of the classical FP determinant~\eref{FPdet}. %
If we now define [cf.~\eref{schrodinger}]
\eq{\label{eq:conditional-wave-2}
\tilde{\psi} &:= |2Vg|^{\frac14}\hat{\mu}\psi\\
&=|2Vg|^{\frac14}\psi+\frac{\I}{4MV}\del{}{x^1}|2Vg|^{\frac14}\psi\\
&=|2Vg|^{\frac14}\psi+\frac{1}{4MV}\hat{\Hs}|2Vg|^{\frac14}\psi+\Omm \ , 
}
and use
\eq{
|2Vg|^{\frac14}\psi &= \tilde{\psi}-\frac{1}{4MV}\hat{\Hs}\tilde{\psi}+\Omm \ ,
}
we can rewrite~\eref{quasi-schrodinger} as
\eq{\label{eq:Heff-schro}
\I\del{\tilde{\psi}}{x^1} = \He\tilde{\psi} \ ,
}
where [cf.~\eref{momentum-operator}]
\begin{align}
\He &:=\hat{\Hs}-\frac{1}{4MV}\hat{\Hs}^2+\frac{1}{2M}\hat{\Pi}_ig^{ij}\hat{\Pi}_j^{\dagger}+\frac{\mathcal{Q}}{M}+\Omm \ ,\label{eq:Heff} \\
\mathcal{Q} &:= |2Vg|^{-\frac14}\del{}{x^1}\left[\del{}{x^1}\left(|2Vg|^{-\frac14}\right)\frac{\sqrt{|2Vg|}}{4V}\right] \ . \label{eq:calQ}
\end{align}
Assuming that $\hat{\Hs}$ is self-adjoint with respect to the standard $L^2$ norm in the $q$-configuration space, we notice that the effective Hamiltonian~\eref{Heff} is a symmetric quantization of~\eref{Hamiltonian-BO} with a quantum correction to the potential given by $\mathcal{Q}$. This correction follows from the choice of the Laplace-Beltrami ordering of the Hamiltonian constraint [cf.~\eref{WDW-BO} and~\eref{nabla2}]. In particular, the adjoint of $\hat{\Pi}_j$ in~\eref{Heff} is taken with respect to the $L^2$ norm that follows from
\eq{\label{eq:L2-IP}
\braket{\Psi_2|\Psi_1}&:= \int\D x\D q\ \tilde{\psi}_2^*(x,q)\tilde{\psi}_1(x,q) \ ,\\
\D x\D q&:= \prod_{i=2}^n\D x^i\prod_{\mu=1}^d\D q^{\mu}\ ,
}
and we assume that the term quadratic in the momenta $\hat{\Pi}_i$ in~\eref{Heff} is not only symmetric but indeed self-adjoint. We thus see from~\eref{Heff-schro},~\eref{Heff} and~\eref{calQ} that the dynamics is unitary up to the NLO because the inner product~\eref{L2-IP} is conserved with respect to $x^1$.\footnote{This settles the issue of perturbative unitarity of the BO weak-coupling expansion. In the literature, there were claims that the expansion leads to terms that violate unitarity~\cite{Kiefer:1991,BKK-0-0,BKK-0-1,BKK-1,BKK-2,BKK-3,Stein2}. As we see in~\eref{Heff-schro} and~\eref{L2-IP}, this is not true because the would-be problematic terms are absorbed into the definition of $\hat{\mu}$ [cf.~\eref{pre-FP-2}] and, ultimately, of the measure [see~\eref{gauge-fixed-IP}]. On the other hand, another set of articles (see, e.g.,~\cite{Bertoni:1996,Bologna:2017,Kiefer:2018}) has claimed that unitarity follows from the inclusion of terms that encode the ``backreaction'' onto the $Q$ fields, which typically involves partial averages over the $q$ fields. This was criticized in~\cite{Chataig:2019-1,Chataig:2021,Chataig:Thesis}, where it was argued that such backreaction is ambiguous due to the ambiguity of the traditional BO factorization $\Psi = \psi_0(Q)\psi(Q;q)$, which was also pointed out in~\cite{Kiefer:2018} (see also Footnote~\ref{foot:traditional}). As was argued in~\cite{Chataig:2021} (see also~\cite{Chataig:Thesis}), the unitarity related to this backreaction, in fact, follows from a wave function redefinition with a non-linear dependence on the choice of state. On the contrary, the construction of~\eref{gauge-fixed-IP} is linear and simply results from the weak-coupling expansion.} Notice that~\eref{L2-IP} is positive definite, and it defines an inner product for the solutions to the WDW constraint~\eref{WDW-BO}. Indeed, using~\eref{BO-ansatz} with $\mathcal{W}_0(Q,q)\equiv\W_0(Q)$ together with~\eref{conditional-wave-2} and the self-adjointness of $\hat{\Hs}$, we can rewrite~\eref{L2-IP} as
\eq{\label{eq:gauge-fixed-IP}
\braket{\Psi_2|\Psi_1}=\int\D x\D q\sqrt{|2Vg|}\ \Psi_2^*(x,q)\,\widehat{|\Delta|}\,\Psi_1(x,q)+\Omm \ ,
}
where\footnote{In the context of quantum optics, a modification of the usual $L^2$ norm was also considered by L\"{a}mmerzahl~\cite{Lammerzahl}.}
\eq{
\widehat{|\Delta|} := 1+\frac{1}{2MV}\hat{\Hs}
}
is a quantization of the (absolute value of the) FP determinant up to the NLO [cf.~\eref{absolute-FP}].

\subsection{Conditional probabilities}
If we evaluate the conserved inner product at an instant $x^1=t$, we can recast~\eref{gauge-fixed-IP} in the gauge-fixed form
\eq{\label{eq:gauge-fixed-IP-2}
\braket{\Psi_2|\Psi_1} =\int\D Q\D q\sqrt{|2Vg|}\,|J|\ \left(\widehat{|\Delta|}^{\frac12}\Psi_2\right)^*\delta(x^1(Q)-t)\,\widehat{|\Delta|}^{\frac12}\Psi_1+\Omm \ ,
}
where the integration now extends over all $Q$ and $q$ variables, and $J$ is the Jacobian $\det\partial x/\partial Q$ associated with the change of coordinates~\eref{normalization-basis-vectors}. The form~\eref{gauge-fixed-IP-2} of the inner product makes it clear that it is an operator version of the Faddeev-Popov gauge-fixing technique~\cite{FP-1,FP-2} for path integrals, in which the gauge condition is fixed by inserting a delta function as well as the FP determinant.\footnote{The definition of the inner product for solutions to the WDW equation by a gauge-fixing procedure has been considered before~\cite{Barvinsky,Woodard:1993,HT:book,Greensite:1989} but its connection with WKB time and the weak-coupling expansion was only elucidated in~\cite{Chataig:2019-1,Chataig:2021,Chataig:Thesis} and the current article to the best of our knowledge. In~\cite{Varda}, the question of equivalence of the Born-Oppenheimer approach and the gauge-fixing method was posed and analyzed for a simple toy model.} 

The probability density related to the inner product~\eref{gauge-fixed-IP-2} reads [cf.~\eref{conditional-wave-2}]
\eq{\label{eq:CP}
p_{\Psi} &:=\left.\frac{|\tilde{\psi}(x,q)|^2}{\braket{\Psi|\Psi}}\right|_{x^1 = t} \\
&=\frac{\sqrt{|2Vg|}}{\braket{\Psi|\Psi}}\left.\left(\widehat{|\Delta|}^{\frac12}\Psi\right)^{*}\widehat{|\Delta|}^{\frac12}\Psi\right|_{x^1 = t}+\Omm \ .
}
To which observations do these probabilities refer? They concern the measurement of the fields $q$ and $x^i$ ($i = 2,\ldots,n$) based on the condition that $x^1$ has (been observed to have) a certain fixed value labeled by $t$. In this way, the probability density~\eref{CP} can be interpreted as a conditional probability density. The fact that solutions to the WDW constraint are used to define conditional probabilities is a symptom of the relational nature of the gauge-fixing: by choosing the WKB time as a clock, one describes $q$ and $x^i$ relative to (or conditioned on) $x^1$.\footnote{The connection between a relational account of quantum theory and the use of conditional probabilities has been discussed before in~\cite{Barvinsky} and, more recently, in~\cite{Hoehn:Trinity, Chataig:2020,Hoehn:2020,Chataig:2021,Chataig:Thesis}, for example. Here, we focus on showing that the Born-Oppenheimer weak-coupling expansion of the WDW constraint leads to an instance of such a relational theory.}

Based on such a relational interpretation, we also suggest that operators defined by the matrix elements
\eq{\label{eq:relObs}
&\braket{\Psi_2|\hat{\Ob}[f|x^1=t]|\Psi_1} \\
&:=\int\D Q\D q\sqrt{|2Vg|}\,|J|\ \left(\widehat{|\Delta|}^{\frac12}\Psi_2\right)^*\delta(x^1(Q)-t)\, \hat{f}\,\widehat{|\Delta|}^{\frac12}\Psi_1+\Omm \ ,
}
which correspond to the insertion of  an operator $\hat{f}$ (which commutes with $x^1$) into~\eref{gauge-fixed-IP-2}, should be seen as relational observables that describe the evolution relative to WKB time.\footnote{See~\cite{Chataig:2019-2,Chataig:2020,Chataig:2021,Chataig:Thesis,Woodard:1985} for preceding treatments.}

\section{Conclusions}
The BO weak-coupling expansion has proved rather useful in phenomenological calculations, especially in the computation of quantum-gravitational corrections to primordial power spectra~\cite{BKK-0-0,BKK-0-1,Bologna:2013,Bologna:2014,Bologna:2016,Bologna:2018,Stein1,Stein2,Mariam,BKK-3,Bologna:2020,BKK-1,BKK-2,Bologna:2017,Chataig:2021}. Nevertheless, its status as a consistent approach to quantum gravity, in particular in regard to its unitarity and its connection with standard gauge fixing techniques, has remained unclear. In this article, we have shown that the perturbative expansion procedure leads to a conserved, positive definite, gauge-fixed inner product, in which the chosen time variable is WKB time. This corresponds to a relational theory, in which the dynamics of all degrees of freedom is described relative to the time parameter that dictates the dynamics of the $Q$ variables (associated to the large scale $M$) when the backreaction of the $q$ variables is neglected. The quantum probabilities may be seen as conditional and are conserved with respect to WKB time.

These results, which were detailed in~\cite{Chataig:Thesis}, pave the way to a unification of the standard BO approach to the problem of time with more recent works that explore the concept of relational observables~\cite{Chataig:2019-2,Hoehn:2018-1,Hoehn:2018-2,Hoehn:2019,Hoehn:Trinity,Chataig:2020,Hoehn:2020,Chataig:Thesis} and the construction of a relational account of quantum gravity. As such a relational depiction can remain valid in a non-perturbative regime, this shows that the phenomenological works based on the weak-coupling expansion (such as~\cite{BKK-0-0,BKK-0-1,Bologna:2013,Bologna:2014,Bologna:2016,Bologna:2018,Stein1,Stein2,Mariam,BKK-3,Bologna:2020,BKK-1,BKK-2,Bologna:2017,Chataig:2021}) could, perhaps, be extended beyond the perturbative regime with a judicious choice of relational time. In fact, even in the regime of validity of perturbation theory in powers of $1/M$, the formalism presented here implies that WKB time remains a valid choice of parametrization, i.e., one can still use the $x^1$ time variable even when incorporating higher orders in the expansion of~\eref{quasi-schrodinger}, such that its implementation need not be restricted to a purely semiclassical regime. In this way, in our quest to a complete theory of quantum gravity, it is possible that we may go beyond semiclassical time.

\section*{Acknowledgements}
The author thanks Claus Kiefer, Manuel Kr\"{a}mer, Branislav Nikoli\'{c}, and David Brizuela for useful discussions over the years; Alexander Y. Kamenshchik, Alessandro Tronconi, and Giovanni Venturi for interesting dialogues; and the Dipartimento di Fisica e Astronomia of the Universit\`{a} di Bologna as well as the I.N.F.N. Sezione di Bologna for financial support.


\begin{thebibliography}{99}

\bibitem{Kuchar:1991}
K.~V.~Kucha\v{r},
\href{https://doi.org/10.1142/S0218271811019347}{Int. J. Mod. Phys. D {\bf 20}, 3} (2011).

\bibitem{Isham:1992}
C.~J.~Isham,
NATO Sci. Ser. C {\bf 409}, 157 (1993).

\bibitem{Kiefer:book}
C.~Kiefer,
{\it Quantum Gravity}, 3rd ed.,
International Series of Monographs on Physics, vol. 155
(Oxford University Press, Oxford, 2012).

\bibitem{Anderson:book}
E.~Anderson,
\href{https://doi.org/10.1007/978-3-319-58848-3}{\it The Problem of Time}, Fundamental Theories of Physics Vol. 190
(Springer, Cham, Switzerland, 2017).

\bibitem{DeWitt:1967}
B.~S.~DeWitt,
\href{https://doi.org/10.1103/PhysRev.160.1113}{Phys. Rev. \textbf{160}, 1113} (1967).

\bibitem{LapRuba:1979}
V.~G.~Lapchinsky and V.~A.~Rubakov,
\href{http://inspirehep.net/record/148275/files/v10p1041.pdf}{Acta Phys. Pol. B {\bf 10}, 1041 (1979)}.

\bibitem{Banks:1984}
T.~Banks,
\href{https://doi.org/10.1016/0550-3213(85)90020-3}{Nucl. Phys. {\bf B 249}, 332 (1985)}.

\bibitem{Halliwell:1984}
J.~J.~Halliwell and S.~W.~Hawking,
\href{https://doi.org/10.1103/PhysRevD.31.1777}{Phys. Rev. D {\bf 31}, 1777 (1985)}.

\bibitem{Brout:1987-4}
R.~Brout,
\href{https://doi.org/10.1007/BF01882790}{Found. Phys. {\bf 17}, 603 (1987)};

R.~Brout and G.~Venturi,
\href{https://doi.org/10.1103/PhysRevD.39.2436}{Phys. Rev. D {\bf 39}, 2436 (1989)}.

\bibitem{PadSingh:1990-1}
T.~P.~Singh and T.~Padmanabhan,
\href{https://doi.org/10.1016/0003-4916(89)90180-2}{Ann. Phys. (N.Y.) {\bf 196}, 296 (1989)}.

\bibitem{PadSingh:1990-2}
T.~Padmanabhan and T.~P.~Singh,
\href{https://doi.org/10.1088/0264-9381/7/3/015}{Class. Quant. Grav. {\bf 7}, 411 (1990)}.

\bibitem{Singh:1990}
T.~P.~Singh,
\href{https://doi.org/10.1088/0264-9381/7/7/006}{Class. Quant. Grav. {\bf 7}, L149 (1990)}.

\bibitem{Kiefer:1991}
C.~Kiefer and T.~P.~Singh,
\href{https://doi.org/10.1103/PhysRevD.44.1067}{Phys. Rev. D {\bf 44}, 1067 (1991)}.

\bibitem{Bertoni:1996}
C.~Bertoni, F.~Finelli and G.~Venturi,
\href{https://doi.org/10.1088/0264-9381/13/9/005}{Class. Quant. Grav. {\bf 13}, 2375 (1996)}.

\bibitem{Kiefer:1993-2}
C.~Kiefer,
\href{https://doi.org/10.1007/3-540-58339-4_19}{\it The semiclassical approximation to quantum gravity},
in: {\it Canonical Gravity: From Classical to Quantum}, Lecture Notes in Physics, vol.~434, edited by J.~Ehlers and H.~Friedrich
(Springer, Berlin, 1994).

\bibitem{Parentani:1997-2}
R.~Brout and R.~Parentani,
\href{https://doi.org/10.1142/S0218271899000031}{Int. J. Mod. Phys. D {\bf 8}, 1 (1999)}.

\bibitem{Bologna:2017}
A.~Y.~Kamenshchik, A.~Tronconi and G.~Venturi,
\href{https://doi.org/10.1088/1361-6382/aa8fb3}{Class. Quant. Grav. {\bf 35}, 015012 (2018)}.

\bibitem{Kiefer:2018}
C.~Kiefer and D.~Wichmann,
\href{https://doi.org/10.1007/s10714-018-2390-4}{Gen. Relativ. Gravit. {\bf 50}, 66 (2018)}.
  
\bibitem{Chataig:2019-1}
L.~Chataignier,
\href{https://doi.org/10.1515/zna-2019-0223}{Z. Naturforsch. A \textbf{74}, 1069 (2019)}.

\bibitem{Rovelli:1990-1}
  C.~Rovelli,
  Phys.\ Rev.\ D \href{https://doi.org/10.1103/PhysRevD.42.2638}{{\bf 42} 2638} (1990).

\bibitem{Rovelli:1990-2}
  C.~Rovelli,
  Classical Quantum Gravity\  \href{https://doi.org/10.1088/0264-9381/8/2/011}{{\bf 8} 297} (1991);
    \href{https://doi.org/10.1088/0264-9381/8/2/012}{{\bf 8} 317} (1991).

\bibitem{Rovelli:1991}
  C.~Rovelli,
  Phys.\ Rev.\ D \href{https://doi.org/10.1103/PhysRevD.43.442}{{\bf 43}, 442} (1991).

\bibitem{Dittrich:2004}
  B.~Dittrich,
  Gen.\ Relativ.\ Gravit.\  \href{https://doi.org/10.1007/s10714-007-0495-2}{{\bf 39} 1891} (2007).

\bibitem{Dittrich:2005}
  B.~Dittrich,
  Classical Quantum Gravity\  \href{https://doi.org/10.1088/0264-9381/23/22/006}{{\bf 23} 6155} (2006).

\bibitem{Tambor}
J.~Tambornino,
\href{https://doi.org/10.3842/SIGMA.2012.017}{SIGMA \textbf{8}, 017} (2012).

\bibitem{Chataig:2019-2}
L.~Chataignier,
\href{https://doi.org/10.1103/PhysRevD.101.086001}{Phys. Rev. D \textbf{101}, 086001 (2020)}.

\bibitem{Hoehn:2018-1}
A.~Vanrietvelde, P.~A.~H\"{o}hn, F.~Giacomini and E.~Castro-Ruiz,
\href{https://doi.org/10.22331/q-2020-01-27-225}{Quantum {\bf 4}, 225 (2020)}.

\bibitem{Hoehn:2018-2}
P.~A.~H\"{o}hn and A.~Vanrietvelde,
 \href{https://doi.org/10.1088/1367-2630/abd1ac}{New J. Phys. {\bf 22}, 123048 (2020)}.

\bibitem{Hoehn:2019}
P.~A.~H\"{o}hn,
\href{https://doi.org/10.3390/universe5050116}{Universe {\bf 5},  116 (2019)}.

\bibitem{Hoehn:Trinity}
P.~A.~H\"{o}hn, A.~R.~H.~Smith and M.~P.~E.~Lock,
\href{https://doi.org/10.1103/PhysRevD.104.066001}{Phys. Rev. D 104, 066001} (2021).

\bibitem{Chataig:2020}
L.~Chataignier,
\href{https://doi.org/10.1103/PhysRevD.103.026013}{Phys. Rev. D {\bf 103}, 026013 (2021)}.

\bibitem{Hoehn:2020}
P.~A.~H\"{o}hn, A.~R.~H.~Smith and M.~P.~E.~Lock,
\href{https://doi.org/10.3389/fphy.2021.587083}{Front. in Phys. \textbf{9}, 181} (2021).

\bibitem{Chataig:Thesis}
L.~Chataignier, \href{https://doi.org/10.1007/978-3-030-94448-3}{\it Timeless Quantum Mechanics and the Early Universe}, Springer Theses (Springer, Cham, Switzerland, 2022).

\bibitem{BO:1927}
M.~Born and R.~Oppenheimer,
\href{https://doi.org/10.1002/andp.19273892002}{Ann. Phys. (Berlin) {\bf 389}, 457 (1927)}.

\bibitem{Cederbaum:2008}
L.~S.~Cederbaum,
\href{https://doi.org/10.1063/1.2895043}{J. Chem. Phys. {\bf 128}, 124101 (2008)}.
  
\bibitem{Abedi:2010}
A.~Abedi, N.~T.~Maitra and E.~K.~U.~Gross,
\href{https://doi.org/10.1103/PhysRevLett.105.123002}{Phys. Rev. Lett. {\bf 105}, 123002 (2010)}.
  
\bibitem{Arce:2012}
J.~C.~Arce,
\href{https://doi.org/10.1103/PhysRevA.85.042108}{Phys. Rev. A {\bf 85}, 042108 (2012)}.

\bibitem{FP-1}
L.~D.~Faddeev and V.~N.~Popov,
\href{https://doi.org/10.1016/0370-2693(67)90067-6}{Phys. Lett. {\bf 25B}, 29 (1967)}.

\bibitem{FP-2} 
L.~D.~Faddeev and V.~N.~Popov,
\href{https://doi.org/10.1070/PU1974v016n06ABEH004089}{Sov. Phys. Usp. {\bf 16}, 777 (1974)}
[Usp.\ Fiz.\ Nauk {\bf 111}, 427 (1973)].

\bibitem{HT:book}
M.~Henneaux and C.~Teitelboim,
{\it Quantization of Gauge Systems}
(Princeton University Press, Princeton, New Jersey, 1992).

\bibitem{BKK-1}
D.~Brizuela, C.~Kiefer and M.~Kr\"{a}mer,
\href{https://doi.org/10.1103/PhysRevD.93.104035}{Phys. Rev. D \textbf{93}, 104035 (2016)}.

\bibitem{BKK-2}
D.~Brizuela, C.~Kiefer and M.~Kr\"{a}mer,
\href{https://doi.org/10.1103/PhysRevD.94.123527}{Phys. Rev. D \textbf{94}, 123527 (2016)}.

\bibitem{BKK-0-0}
C.~Kiefer and M.~Kr\"{a}mer,
\href{https://doi.org/10.1103/PhysRevLett.108.021301}{Phys. Rev. Lett. \textbf{108}, 021301 (2012)}.

\bibitem{BKK-0-1}
D.~Bini, G.~Esposito, C.~Kiefer, M.~Kr\"{a}mer and F.~Pessina,
\href{https://doi.org/10.1103/PhysRevD.87.104008}{Phys. Rev. D \textbf{87}, 104008 (2013)}.

\bibitem{Bologna:2013}
 A.~Y.~Kamenshchik, A.~Tronconi and G.~Venturi,
\href{https://doi.org/10.1016/j.physletb.2013.08.067}{Phys. Lett. B {\bf 726}, 518 (2013)}.

\bibitem{Bologna:2014}
A.~Y.~Kamenshchik, A.~Tronconi and G.~Venturi,
\href{https://doi.org/10.1016/j.physletb.2014.05.028}{Phys. Lett. B {\bf 734}, 72 (2014)}.

\bibitem{Bologna:2016}
A.~Y.~Kamenshchik, A.~Tronconi and G.~Venturi,
\href{https://doi.org/10.1103/PhysRevD.94.123524}{Phys. Rev. D {\bf 94}, 123524 (2016)}.

\bibitem{Bologna:2018}
A.~Tronconi, G.~P.~Vacca and G.~Venturi,
\href{https://doi.org/10.1103/PhysRevD.67.063517}{Phys. Rev. D \textbf{67} 063517 (2003)};
A.~Y.~Kamenshchik, A.~Tronconi, T.~Vardanyan and G.~Venturi,
\href{https://doi.org/10.1103/PhysRevD.97.123517}{Phys. Rev. D {\bf 97}, 123517 (2018)}.

\bibitem{Stein1}
C.~F.~Steinwachs and M.~L.~van der Wild,
\href{https://doi.org/10.1088/1361-6382/aac587}{Class. Quant. Grav. \textbf{35}, 135010 (2018)}.

\bibitem{Stein2}
C.~F.~Steinwachs and M.~L.~van der Wild,
\href{https://doi.org/10.1088/1361-6382/ab3a1b}{Class. Quant. Grav. \textbf{36}, 245015 (2019)}.

\bibitem{Mariam}
M.~Bouhmadi-L\'{o}pez, M.~Kr\"{a}mer, J.~Morais and S.~Robles-P\'{e}rez,
\href{https://doi.org/10.1088/1475-7516/2019/02/057}{J. Cosmol. Astropart. Phys. 02 (2019) 057}.

\bibitem{BKK-3}
D.~Brizuela, C.~Kiefer, M.~Kr\"{a}mer and S.~Robles-P\'{e}rez,
\href{https://doi.org/10.1103/PhysRevD.99.104007}{Phys. Rev. D \textbf{99}, 104007 (2019)}.

\bibitem{Bologna:2020}
A.~Y.~Kamenshchik, A.~Tronconi and G.~Venturi,
\href{https://arxiv.org/abs/2010.15628}{arXiv:2010.15628}.

\bibitem{Chataig:2021}
L.~Chataignier and M.~Kr\"amer,
\href{https://doi.org/10.1103/PhysRevD.103.066005}{Phys. Rev. D \textbf{103}, 066005} (2021).

\bibitem{Zeh}
H.~D.~Zeh,
\href{https://doi.org/10.1016/0375-9601(88)90842-0}{Phys. Lett. A {\bf 126}, 311} (1988).

\bibitem{Kiefer:topology}
C.~Kiefer,
\href{https://doi.org/10.1103/PhysRevD.47.5414}{Phys. Rev. D \textbf{47}, 5414 (1993)}.

\bibitem{DeWitt:1957}
B.~S.~DeWitt,
\href{https://doi.org/10.1103/RevModPhys.29.377}{Rev. Mod. Phys. \textbf{29}, 377 (1957)}.

\bibitem{Barvinsky}
A.~Barvinsky,
\href{https://doi.org/10.1016/0370-1573(93)90032-9}{Phys. Rep. \textbf{230}, 237 (1993)}.

\bibitem{Woodard:1993}
R.~P.~Woodard,
\href{https://doi.org/10.1088/0264-9381/10/3/008}{Class. Quant. Grav. {\bf 10}, 483 (1993)}.

\bibitem{Greensite:1989}
J.~Greensite, \href{https://doi.org/10.1016/0550-3213(90)90196-K}{Nucl. Phys. {\bf B342} 409 (1990)}.

\bibitem{Varda}
A.~Y.~Kamenshchik, A.~Tronconi, T.~Vardanyan and G.~Venturi,
\href{https://doi.org/10.1142/S0218271819500731}{Int. J. Mod. Phys. D \textbf{28}, 1950073} (2019).

\bibitem{Lammerzahl}
C.~L\"{a}mmerzahl,
\href{https://doi.org/10.1016/0375-9601(95)00345-4}{Phys. Lett. A {\bf 203}, 12 (1995)}.

\bibitem{Woodard:1985}
N.~C.~Tsamis and R.~P.~Woodard,
\href{https://doi.org/10.1088/0264-9381/2/6/011}{Class. Quant. Grav. {\bf 2}, 841 (1985)}.

\end{thebibliography}
\end{document}